\documentclass{elsart}

\usepackage{epsfig}

\usepackage{amssymb}
\makeatletter
\newcounter{subeqncnt}
\def\thesubeqncnt{\alph{subeqncnt}}
\def\subequations{\begingroup%
\stepcounter{equation}\edef\@tempa{\theequation}%
\let\c@equation\c@subeqncnt\c@subeqncnt\z@
 \edef\theequation{\@tempa\noexpand\thesubeqncnt}}
  
  \makeatother

\begin{document}

\begin{frontmatter}


\title{Measurement of $K^+ \rightarrow \pi^0 \mu^+ \nu \gamma$
decay using stopped kaons}

\thanks[email]{ Corresponding author.\\{\it E-mail address:}
 suguru@phys.sci.osaka-u.ac.jp~(Suguru SHIMIZU)}

\author[osa]{S.~Shimizu \thanksref{email}}
\author[osa]{K.~Horie \thanksref{kate}},
\author[inr]{M.A.~Aliev}, 
\author[tuku1]{Y.~Asano}, 
\author[kek]{T.~Baker}, 
\author[dep]{P.~Depommier}, 
\author[has]{M.~Hasinoff}, 
\author[kek]{Y.~Igarashi},
\author[kek]{J.~Imazato}, 
\author[inr]{A.P.~Ivashkin}, 
\author[inr]{M.M.~Khabibullin}, 
\author[inr]{A.N.~Khotjantsev},
\author[inr]{Y.G.~Kudenko},
\author[inr]{A.S. Levchenko},
\author[kek]{G.Y.~Lim}, 
\author[tri]{J.A.~Macdonald \thanksref{jam}}, 
\author[inr]{O.V.~Mineev}, 
\author[sas]{C.~Rangacharyulu},
and
\author[kek]{S.~Sawada}

\collab{KEK-E470 Collaboration}
\address[osa]{Department of Physics, Osaka University, Osaka 560-0043,
 Japan }
\address[inr]{
Institute for Nuclear Research,
Russian Academy of Sciences, Moscow 117312, 
      Russia }
\address[tuku1]{
 Institute of Applied Physics, University of Tsukuba, Ibaraki 305-0006, 
 Japan }
\address[kek]{
 Institute of Particle and Nuclear Studies (IPNS), High Energy Accelerator Research Organization (KEK), Ibaraki  305-0801, Japan }
\address[dep]{
Laboratoire de Physique Nucl\'eaire, Universit\'e de Montr\'{e}al,
Montr\'{e}al, Qu\'ebec, Canada H3C 3J7}
\address[has]{
 Department of Physics and Astronomy, University of British Columbia, 
 Vancouver, Canada V6T 1Z1 }
\address[tri]{
TRIUMF, Vancouver, British Columbia, Canada V6T 2A3 }
\address[sas]{
Department of Physics, University of Saskatchewan, Saskatoon, Canada 
S7N 5E2}

\thanks[kate]{ Present address:  Research Center for Nuclear Physics
(RCNP), Osaka University, Ibaraki, Osaka 567-0043, Japan}
\thanks[jam]{Deceased}

\begin{abstract}
The $K^+ \rightarrow \pi^0 \mu^+ \nu \gamma$ ($K_{\mu 3 \gamma}$)
decay has been measured with stopped positive kaons at the KEK 12 GeV
proton synchrotron. A $K_{\mu 3 \gamma}$ sample containing 125 events
was obtained. The partial branching ratio
$Br(K_{\mu 3 \gamma},  E_{\gamma}>30 {\rm~ MeV}, \theta_{\mu^+
\gamma}>20^{\circ})$ was found to be $[2.4 \pm 0.5(stat) \pm
0.6(syst)]\times 10^{-5}$, which is in good agreement with
theoretical predictions. 
\end{abstract}

\begin{keyword}
Kaon decays
\PACS 13.20.Eb \sep 14.40.Aq
\end{keyword}

\end{frontmatter}

Semi-leptonic radiative decays of K-mesons, $K \rightarrow \pi l \nu
\gamma$ ($K_{l3\gamma}$), offer a
good testing ground of hadron structure models making use of
low-energy effective Lagrangians
inspired by chiral perturbation theory (ChPT).   It
is expected that branching ratio measurements and decay spectra
with a single pion in the final state provide simple but good
constraints on the models.  The radiative decays of mesons usually
consist of an internal bremsstrahlung (IB) process and a
hadron-structure-dependent direct emission (DE) process.  While the IB
process is dominant in the decays with electrons in the final state
such as $K \rightarrow  \pi e \nu \gamma$ ($K_{e3\gamma}$) decays,
one expects a significant DE contribution when there is a muon in the final
state, $K \rightarrow \pi \mu \nu \gamma$ ($K_{\mu 3 \gamma}$),
because of the larger lepton mass. The relative size of the DE effects 
can be calculated in strong interaction models. Following the
early estimates~\cite{fis69,fea70-1,fea70-2}, based on current
algebra, calculations in the framework of the ChPT theory have been
done~\cite{bij93}.  

$K_{l3 \gamma}$ branching ratios  of neutral kaons with
electrons and muons in the final state have been reported in the
literature with branching fractions of  $K_{e3\gamma}^0$ and 
$K_{\mu3\gamma}^0$ of $3.5 \times 10^{-3}$ and $5.5 \times  10^{-4}$,
respectively~\cite{pdb02}. For the charged kaons, a $K_{e3\gamma}^+$
decay branching ratio of $2.65 \times 10^{-4}$ has been
measured~\cite{pdb02}, and results of the first measurement of the
$K_{\mu 3 \gamma}^-$ decay using an in-flight $K^-$ beam has recently
been reported~\cite{tch05}. In this paper, we present a new
measurement of the $K^+ \rightarrow \pi^0 \mu^+ \nu \gamma$ $(K_{\mu 3
\gamma}^+$) decay using a stopped $K^+$ 
beam along with detailed Monte Carlo simulations, which enabled us to
determine the $K^+_{\mu3 \gamma}$ branching ratio.    

The experiment was performed at the KEK 12-GeV proton synchrotron.
The detector was basically the E246 setup~\cite{main}, which had
the 12-sector toroidal spectrometer and the ancillary detector
assemblies such as the  
photon calorimeter and the particle tracking system.   Since the 
system was built primarily for the purpose of a high precision
test of time reversal invariance in the  
$K^+ \rightarrow \pi^0 \mu^+ \nu$ ($K_{\mu 3}$) 
decay~\cite{main}, an elaborate simulation program based on
GEANT3~\cite{geant} has been developed.  Details of the setup are well
documented in Ref.~\cite{nim}. In addition to the $T$-violation search,
spectroscopic studies for various decay channels have also been
successfully performed using the same detector
system~\cite{shi00,kpi2gpap,ke4pap}.

A separated 660-MeV/$c$ $K^+$ beam was stopped in an active target
system. The $K_{\mu 3 \gamma}$ events were identified by analyzing the
$\mu^+$  momentum with the spectrometer and detecting three photons in
the CsI(Tl) calorimeter. The momentum vectors of the charged particles were
determined by reconstructing their trajectories  in the spectrometer
using multi-wire proportional chambers (MWPCs). The $\mu^+$s were
selected by determining the squared  mass ($M_{\rm TOF}^2$) from a 
time-of-flight measurement. The photon energy and hit position were
obtained, respectively, by summing the energy deposits and taking the
energy-weighted centroid of the CsI(Tl) crystals sharing a shower. The
analysis procedures of the present work for the charged particle
tracking, TOF measurement, and photon energy and hit position
determinations are the same as those of the previous $K_{\pi 2
\gamma}$ study (see Ref.~\cite{kpi2gpap} for details).

\begin{figure}
\includegraphics{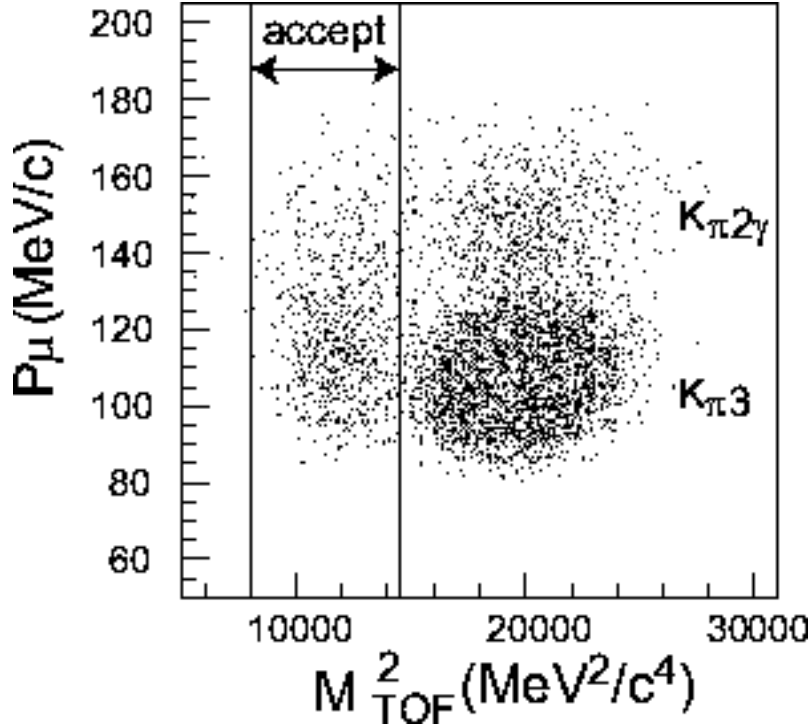}
\caption{Correlation plot of $M_{\rm TOF}^2$ and $P_{\mu}$. 
The $K_{\pi 3}$ and $K_{\pi 2 \gamma}$ events which were used to
calculate the $K_{\mu 3 \gamma}$ branching ratio and the background
fractions are also seen.
}
\label{evsele}
\end{figure}

\begin{figure}
\scalebox{0.6}{\includegraphics{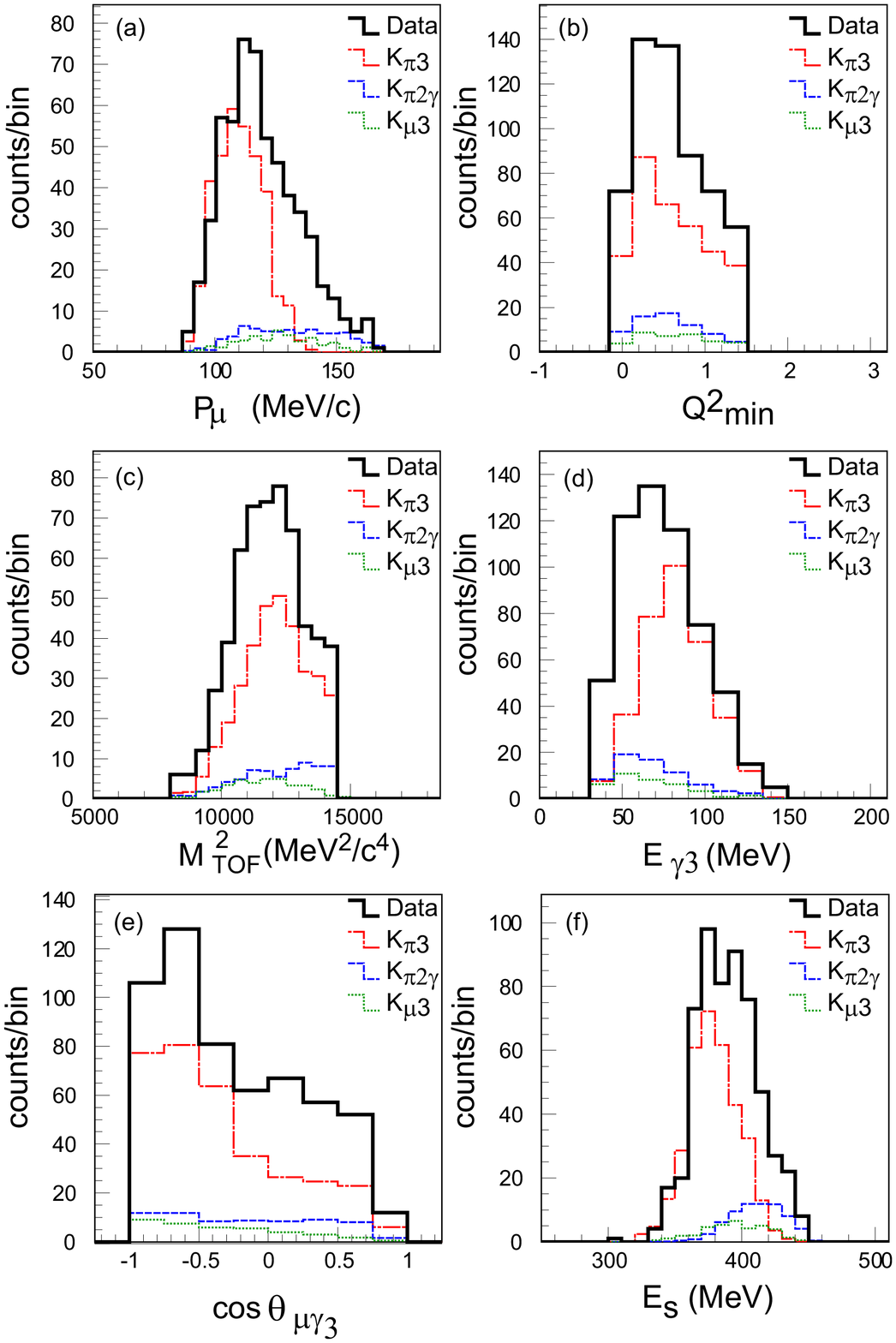}}
\caption{Black(solid) histograms are experimental data extracted with the
present selection conditions for (a) $P_{\mu^+}$, (b) $Q^2$,
(c)$M_{\rm TOF}^2$, (d) $E_{\gamma 3}$, (e) cos$\theta_{\mu^+\gamma
3}$, and (f) $E_s$. Red(dot-dashed), blue(dashed), and green(dot)
histograms are backgrounds due to $K_{\pi3 }$, $K_{\pi 2 \gamma}$, and
$K_{\mu 3}$ decays, respectively.
}
\label{fg:sum1}
\end{figure}

Specific cuts for the $K_{\mu 3 \gamma}$ selection are described
below. The charged particle momentum corrected for the energy loss 
in the target ($P_{\mu^+}$) was imposed to be $P_{\mu^+}<170$ MeV/$c$.
Events from $\pi^+$ decays in-flight and scattering of the charged
particle from the magnet pole faces were eliminated by requiring the
particle track to be consistent with the hit position in the ring
counters surrounding the active target system~\cite{nim}. The
selection criterion for muons was $8000<M_{\rm TOF}^2<14500$
MeV$^2/c^4$, as shown in Fig.~\ref{evsele}. Events with three photon
clusters in the calorimeter were selected: two as coming from $\pi^0
\rightarrow \gamma_1 \gamma_2$ and one being a radiative photon
($\gamma_3$). Since there are three possible combinations to form a
$\pi^0$ from three photons, a quantity $Q^2$ was introduced to find 
the correct pairing,  
\begin{eqnarray}
        Q^2=(M_{\pi^0}-M)^2/\sigma_{M}^2 +({\rm cos}\theta_{\mu^+
        \gamma_3}^{\rm MEA}- {\rm cos}\theta_{\mu^+ \gamma_3}^{\rm CAL}
-\alpha)^2/\sigma_{\alpha}^2, \label{q2}
\end{eqnarray}
where $M_{\pi^0}$ is the invariant mass of
the selected pair and $\theta_{\mu^+ \gamma 3}$ is the opening angle
between the $\mu^+$ and $\gamma_3$. The superscripts  MEA and CAL stand
for the measured angle and the angle calculated  from other
observables by assuming the $K_{\mu 3 \gamma}$ kinematics. The pair
with the minimum $Q^2$ (=$Q^2_{\rm min}$) among the three possible
combinations was adopted as the correct 
pairing. The $\sigma$ ($\sigma_M$, $\sigma_{\alpha}$) and offset values
($M$, $\alpha$) in each terms are $\sigma_M=10.92$ MeV/$c^2$,
$\sigma_{\alpha}=0.273$, $M=118.3$ MeV/$c^2$, and $\alpha =
0.265$. The choice of the parameters were determined to obtain the
highest probability for the  correct pairing by 
using the simulation data. The correct pairing probability was
estimated to be 69\% from the Monte Carlo simulation. Further, since
most of background events do not satisfy the $K_{\mu 3 
\gamma}$ kinematics, the cut of $Q^2_{\rm min}<1.5$ reduced the
background contaminations. An additional
cut condition, cos$[\theta_{\gamma \gamma}]_{\rm min}<0.45$, was
applied to reject events with a photon split into multiple clusters,
where $[\theta_{\gamma \gamma}]_{\rm min}$ is the minimum opening
angle of photons among the three combinations.  The above conditions
were sufficient to select $K_{\mu 3 \gamma}$ events. From this
analysis, a sample with 565 events was extracted. The spectra are
shown in Fig.~\ref{fg:sum1}. The black(solid) histograms are the data
comprising of the $K_{\mu 3 \gamma}$ events and the background events
to be discussed below.

\begin{figure}
\scalebox{0.6}{\includegraphics{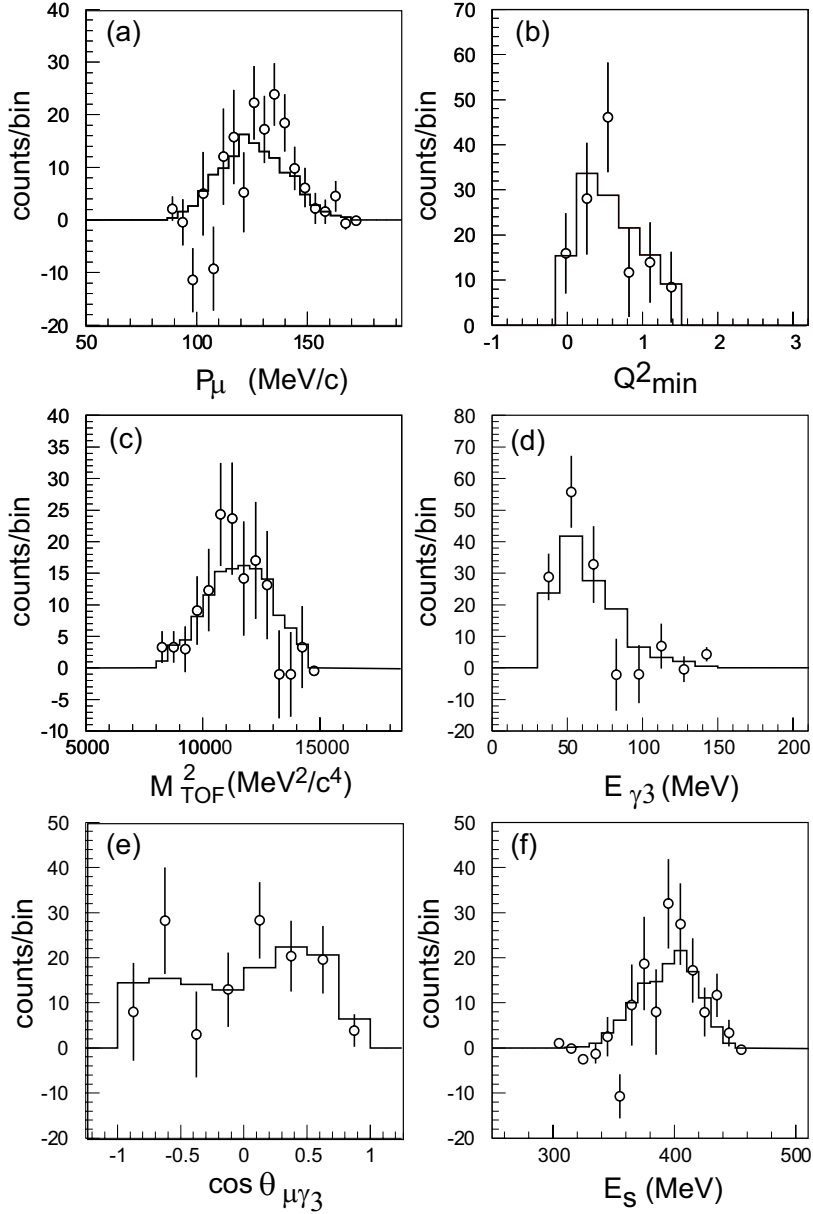}}
\caption{$K_{\mu3 \gamma}$ spectra obtained by subtracting
backgrounds (dots): (a) $P_{\mu^+}$, (b) $Q^2$,
(c)$M_{\rm TOF}^2$, (d) $E_{\gamma 3}$, (e) cos$\theta_{\mu^+
\gamma 3}$, and (f) $E_s$. The histograms are the Monte Carlo simulations.
}
\label{fg:sub1}
\end{figure}

There are three major background components, $K^+ \rightarrow \pi^+
\pi^0 \pi^0$ ($K_{\pi 3}$), $K^+ \rightarrow \pi^+ \pi^0 \gamma$
($K_{\pi 2 \gamma}$), and $K_{\mu 3}$. The former two could imitate 
$K_{\mu 3 \gamma}$ if the pion decays in flight and the three photons
hit the calorimeter. Also, $K_{\mu 3}$ with an accidental photon could
contribute to $K_{\mu 3 \gamma}$. The $K_{\pi3}$ and $K_{\pi 2 \gamma}$
contaminations were estimated using a Monte Carlo simulation. The
simulation data were analyzed in the same manner as the experimental
data, yielding surviving background fractions. In order to determine
these fractions, the results of careful evaluations, carried out in
the previous  $K_{\pi 2 \gamma}$ study~\cite{kpi2gpap}, were
used. These $K_{\pi 3}$ and $K_{\pi 2  \gamma}$ events can be seen in
the $P_{\mu}$-$M_{\rm TOF}^2$ scatter plot in Fig.~\ref{evsele}. The
numbers of the $K_{\pi 3}$ and $K_{\pi2 \gamma}$ events were
calculated from those of the experimental $K_{\pi 3}$ and $K_{\pi 
2 \gamma}$ events by using acceptance ratios as,
\begin{subequations}     
\begin{eqnarray}
	Y(K_{\pi3 }^{BG})= \frac{\Omega(K_{\pi 3}^{BG})}{\Omega(K_{\pi
3}^{NM})} \cdot Y(K_{\pi 3}^{NM}), \\
	Y(K_{\pi2 \gamma}^{BG})= \frac{\Omega(K_{\pi 2
\gamma}^{BG})}{\Omega(K_{\pi 2 \gamma}^{NM})} \cdot Y(K_{\pi2
\gamma}^{NM}), 
\end{eqnarray}
\end{subequations}     
\noindent
where $Y(X)$ is the yield of decay channel $X$ and $\Omega(X)$ is the
detector acceptance determined by the simulation.
$BG$ and $NM$ stand for the selection conditions of the present
background evaluation and the previous normal $K_{\pi 2 \gamma}$
study~\cite{kpi2gpap}, respectively. Potential systematic errors from
the uncertainty of the calculated acceptances can be reduced by taking
the ratio of the acceptances.  Substituting the geometrical acceptance
of our setup, $\Omega(K_{\pi 3}^{BG})=9.91\times 10^{-7}$, $\Omega(K_{\pi
3}^{NM})=9.45\times10^{-5}$, $\Omega(K_{\pi 2\gamma}^{BG})$=$3.61\times
10^{-6}$, and $\Omega(K_{\pi 2\gamma}^{NM})=2.40\times 10^{-4}$ and
the measured yields $Y(K_{\pi3}^{NM})$=32105 and $Y(K_{\pi2
\gamma}^{NM})=4434$, $Y(K_{\pi 3}^{BG})$ and $Y(K_{\pi 2
\gamma}^{BG})$ were determined to be 337 and 67, respectively,
corresponding to 60\% 
and 12\% of the total events. The contribution from $K_{\mu3 }$ was
studied by accepting accidental events in the CsI(Tl) TDC data and its
number was found to be 36. The background spectra are shown in
Fig~\ref{fg:sum1} as the red(dot-dashed), blue(dashed), and green(dot)
histograms for $K_{\pi 3}$, $K_{\pi 2 \gamma}$, and $K_{\mu 3}$,
respectively. Fig.~\ref{fg:sub1} shows the background-subtracted
distributions compared with the simulation described in the
following. The number of $K_{\mu 3 \gamma}$ events was found to be
$125\pm 25$. 

In the simulation, the $K_{\mu 3 \gamma}$ data were generated according to
the matrix elements given in Ref.~\cite{fis69}. Here, only the IB process was
taken into account because the present result is not sensitive enough
to study the DE contribution (see below). The detector acceptance for
$K_{\mu 3 \gamma }$ in the region of $E_{\gamma_3}>30 {\rm~MeV},
\theta_{\mu^+ \gamma_3}>20^{\circ}$ was determined to be
$\Omega(K_{\mu 3 \gamma} )=1.52 \times 10^{-4}$. The reduced $\chi^2$
values of fits between the experimental data and the simulation
are (a)1.33, (b)1.09, (c)0.80, (d)1.24, (e)0.94, and
(f)1.05 in each spectrum, lending credence to our claim that these are
indeed $K_{\mu 3 \gamma}$ events.  

The partial branching ratio of $K_{\mu 3 \gamma}$ in the region of
$E_{\gamma_3}>30 {\rm~MeV}, \theta_{\mu^+ \gamma_3}>20^{\circ}$ can be
derived from separately comparing with the $K_{\pi 2 \gamma}$ and
$K_{\pi3}$ branching ratio by means of the yield and acceptance ratios as, 
\begin{subequations}         
\begin{eqnarray}
	Br(K_{\mu 3 \gamma})&=&\frac{Y(K_{\mu 3 \gamma})}{Y(K_{\pi 2
\gamma}^{NM})}\frac{\Omega(K_{\pi 2 \gamma}^{NM})}{\Omega(K_{\mu 3 \gamma})} \cdot
Br(K_{\pi 2 \gamma}), \label{br1} \\ 
	Br(K_{\mu 3 \gamma})&=&\frac{Y(K_{\mu 3 \gamma})}{Y(K_{\pi
3}^{NM})}\frac{\Omega(K_{\pi3 }^{NM})}{\Omega(K_{\mu 3 \gamma})} \cdot 
Br(K_{\pi 3}), \label{br2}
\end{eqnarray}
\end{subequations}     
where $Br(X)$ is the branching ratio for the process X. Making use of 
$Br(K_{\pi 2 \gamma})=2.61\times 10^{-4}$ predicted by theoretical
calculation~\cite{daf92} and $Br(K_{\pi 3})=1.73\times 10^{-2}$
by the Particle Data Group~\cite{pdb02}, the $K_{\mu 3 \gamma}$
branching ratio was determined to be $Br(K_{\mu 3 \gamma})=(2.4\pm
0.5) \times 10^{-5}$ from $K_{\pi 2 \gamma}$ and $Br(K_{\mu 3
\gamma})=(2.8 \pm 0.6 )\times 10^{-5}$ from $K_{\pi 3}$, which are
consistent with each other.
Since only $\pi^+$s in the
endpoint region of the $K_{\pi 3}$ decays were accepted by the
toroidal spectrometer and the spectrometer acceptance was strongly
affected by the $K_{\pi 3}$ form factors assumed in the simulation, we 
adopted $Br(K_{\mu 3\gamma})$ from Eq.(\ref{br1}) and used
Eq.(\ref{br2}) only for a consistency check of the deduced results.

The major systematic errors in the determination of the $K_{\mu 3
\gamma}$ branching ratio come from uncertainty of the $K_{\pi 3}$ and
$K_{\pi 2 \gamma}$ background fractions. Since we rely on the Monte
Carlo simulation for the background estimation, imperfect
reproducibility of the experimental conditions introduces 
systematic errors. We carefully evaluated the effects from the TOF
measurement and the charged particle tracking, which would distort the
$K_{\mu 3 \gamma}$, $K_{\pi 3}$, $K_{\pi 2 \gamma}$ spectra and, as a
consequence, introduce the systematic errors. These errors were
estimated by varying the inputs of the TOF resolution and the MWPC
spatial resolution in the simulation within their maximum likely
uncertainties (see below). In order to estimate the other systematic
effects, the $K_{\pi 3}$ and $K_{\pi 2 \gamma}$ fractions could be
controlled by requiring an extra cut at the cost of good $K_{\mu 3
\gamma}$ events as follows.   

Since events with $\pi^+$ decays in the spectrometer (DIS) are
seen as a tail under the $\pi^+$ peak, as shown in Fig.~\ref{fg:sum1}(c),
a wrong estimate of the TOF resolution could introduce a systematic
error. This effect was estimated by changing the TOF resolution
assumed in the simulation from $\sigma_{\rm TOF}=270$ ps, which was
experimentally determined using the $K_{e3}$ and $K_{\mu 3}$ decays, to
$\sigma_{\rm TOF}=270\pm 20$ ps. Its contribution to the branching
ratio error was found to be $ \Delta Br(K_{\mu
3 \gamma})=0.1\times 10^{-5}$.  Also, the muon mass cut point
dependence was investigated to study its contribution to the DIS
component and found to be very small. We note that an asymmetric
structure due to the DIS contribution shown in Fig.~\ref{fg:sum1}(c)
was removed by subtracting the $K_{\pi 3}$ and $K_{\pi 2 \gamma}$
backgrounds.

Since most of DIS events were rejected by the consistency cut between
the hit position in the ring counter and the charged particle track, the 
MWPC spatial resolution for the input to the simulation affects the
estimation of the background fraction. In the simulation, we varied
the resolutions by $\pm$10\% from the experimental values to
estimate this effect. The background fraction was obtained by
repeating the same analysis. We arrived at $\Delta Br(K_{\mu3
\gamma})=0.2\times 10^{-5}$  as the contribution to systematic error
from particle tracking.

The background contribution is also influenced by the total energy sum
of the $\mu^+$ and the photons defined as
$E_s=E_{\mu^+}+\sum_{i=1}^{3} E_{\gamma i}$ because the $E_s$
distribution has a specific structure corresponding to the decay
channel, as shown in Fig.~\ref{fg:sum1}(f). Selecting events with the 
conditions of $E_s>400$ MeV and $E_s<400$ MeV, the $K_{\mu 3 \gamma}$
branching ratio was determined to be $(2.9\pm0.6)\times 10^{-5}$ and
$(2.0\pm0.7)\times 10^{-5}$, respectively. The background fractions
using these cut conditions are summarized in Table \ref{tb:bgcon1}. 
It is noted that the $E_s$ variation affects the background 
contributions significantly but does not change the $K_{\mu 3 \gamma}$ 
branching ratio which depends on the ratio of events and acceptances.  
This feature indicates the correct estimation of the $K_{\pi 3 }$ and
$K_{\pi 2 \gamma}$ background fractions and the detector
acceptances. Although these parameters  are consistent within errors,
the parameter shifts, $\Delta Br(K_{\mu3 \gamma})=0.5\times 10^{-5}$,
were treated as a systematic error due to an acceptance uncertainty
for the background events. 

The $\gamma$ mispairing effect was studied by varying the pairing
probability. In addition to the standard $K_{\mu 3 \gamma}$ selection
conditions, an extra cut, cos$\theta_{\gamma1 \gamma2}>-0.5$, was
imposed, where $\theta _{\gamma_1 \gamma_2 }$ is opening angle between
$\gamma_1$ and $\gamma_2$, and the pairing probability became 81\%
with a loss of good $K_{\mu 3 \gamma}$ events,.  The $Br(K_{\mu
3\gamma})$ was found to be  consistent within $0.2\times 10^{-5}$
which was regarded as the systematic error. The statistical
fluctuation of the backgrounds from $K_{\mu 3}$ with an accidental
photon was treated as a systematic error of $0.05\times 10^{-5}$. The
total systematic error was evaluated by adding all these contributions
in quadrature, as shown in Table \ref{tb:sys}. The most dominant
effect is due to the error of the background fractions from the
uncertainty of the detector acceptance. 

In conclusion, we have performed a measurement of the
radiative $K_{\mu 3 \gamma}$ decay using stopped positive kaons. The
data sample of 125 events was obtained by subtracting the $K_{\pi 3}$,
$K_{\pi 2 \gamma}$, and $K_{\mu3}$ backgrounds. The final result for
the branching ratio is $Br(K_{\mu 3 \gamma}, E_{\gamma_3}>30 {\rm MeV},
\theta_{\mu^+ \gamma_3}>20^{\circ})=[2.4\pm
0.5(stat)\pm0.6(syst)]\times 10^{-5}$, by 
normalizing to the theoretical value of internal bremsstrahlung in
the $K_{\pi 2 \gamma}$ decay. This result is in agreement with the
prediction from chiral perturbation theory~\cite{bij93}: 
$Br(K_{\mu 3 \gamma}, E_{\gamma_3}>30 {\rm MeV}, \theta_{\mu
\gamma_3}>20^{\circ})=2.0\times 10^{-5}$. 
The current experimental result  is not yet sensitive enough to
study the contribution from hadron structure
effects, which is predicted to be at the level of a few \% of the
rate~\cite{bij93}, and/or  to search for $T$-violating triple
correlations such  as $P_T=\vec{p_{\gamma}} \cdot (\vec{p_{\mu}} \times
\vec{p_{\pi^0}})$~\cite{fis69,bra03}.  It is conceivable 
that such experiments may be attempted at the new J-PARC facility currently under construction~\cite{j-parc}.

This work has been supported in Japan by a Grant-in-Aid from the Ministry 
of Education, Culture, Sports, Science and Technology, and by JSPS; in 
Russia by the Ministry of Science and Technology, and by the Russian 
Foundation for Basic Research; in Canada by NSERC and IPP, and by the
TRIUMF infrastructure support provided under its NRC contribution. The
authors gratefully acknowledge the excellent support received from the
KEK staff.

\newpage

\listoffigures

\newpage
\begin{table}
\caption{Variation of the $K_{\pi3}$ and $K_{\pi 2\gamma}$ background
fractions with the $E_s$ cuts. The $Y(K_{\mu 3 \gamma})$ and $Br(K_{\mu
3 \gamma})$ values are also shown.  
}  
\label{tb:bgcon1} 
\begin{center} 
\begin{tabular}{crrcc} 
\hline 
$E_s$(MeV) & $K_{\pi 3}$  & $K_{\pi 2 \gamma}$ & $Y(K_{\mu 3 \gamma})$ &
$Br(K_{\mu 3 \gamma}$)$ \times 10^{5}$\\ \hline
no cut& 60\%  & 12\% & 125& $2.4\pm0.5$\\
$E_s>400$& 28\%  & 27\%  & 68 & $2.9\pm0.6$\\
$E_s<400$& 74\%  & 5\% &57 &$2.0\pm0.7$\\ \hline
\end{tabular}
\end{center} 
\end{table}

\begin{table}
\caption{Summary of major systematic errors. All items are added in
quadrature to get the total.}  
\label{tb:sys} 
\begin{center} 
\begin{tabular}{cc} 
\hline
Error source & Uncertainty of $Br(K_{\mu3 \gamma}$) $\times 10^{5}$\\ \hline
TOF resolution& 0.1\\
MWPC resolution& 0.2\\
Detector acceptance  & 0.5\\
$\gamma$ mispairing& 0.2\\
Accidental background& 0.05\\ \hline
Total systematic error& 0.6 \\ \hline
\end{tabular}
\end{center} 
\end{table}


\newpage
\begin{thebibliography}{00}
\bibitem{fis69} E.~Fischbach and J.~Smith, Phys. Rev. {\bf 184} 1645
(1969).
\bibitem{fea70-1} H.~W.~Fearing, E Fischbach, and J. Smith,
Phys. Rev. {\bf D 2} 542 (1970).
\bibitem{fea70-2} H.~W.~Fearing, E Fischbach, and J. Smith,
Phys. Rev. Lett. {\bf 24} 189 (1970).
\bibitem{bij93} J.~Bijinens, G.~Ecker, and J.~Gasser,
Nucl. Phys. {\bf B 396} 81 (1993). 
\bibitem{pdb02} Particle Data Group, Phys. Lett. {\bf B592} 1 
(2004).
\bibitem{tch05} O.~Tchikilev {\em et al.}, hep-ex/0506023 (2005).
\bibitem{main} M.~Abe {\em et al.}, Phys. Rev. Lett. {\bf 83} 4253
(1999);~M.~Abe {\em et al.} Phys. Rev. Lett. {\bf 93} 131601 (2004).
\bibitem{geant}R. Brun {\em et al.}, CERN Program Library Long Writeup
W5013, CERN Applications Software Group (1993). 
\bibitem{nim} J.A.~Macdonald {\em et al.}, Nucl. Instr. and Meth. {\bf
A506} 60 (2003).
\bibitem{shi00} S.~Shimizu {\em et al.}, Phys. Lett. {\bf B495} 33
        (2000);~A.S.~Levchenko {\em et al.}, Phys. At. Nucl. {\bf 65},
	2232 (2002);~K.~Horie {\em et al.}, Phys. Lett. {\bf B513} 311
	(2001);~Y.-H.~Shin {\em et al.}, Eur. Phys. J. {\bf C12} 627 (2000). 
\bibitem{kpi2gpap}M.A.~Aliev {\em et al.}, Phys. Lett. {\bf B554} 7
(2003).
\bibitem{ke4pap}~S.~Shimizu {\em et al.}, Phys. Rev. {\bf D70} 037101 (2004). 
\bibitem{daf92} G.~D'Ambrosio, M.~Miragliuolo, and P.~Santorelli, in
        $Da\Phi ne Physics Handbook$, edited by L.~Maiani, G.~Pancheri,
        and N.~Paver (Laboratori Nazionali di Frascati, Frascati, 1992).
\bibitem{bra03} V.~V.~Braguta, A.~A.~Likhoded, and A.~E.~Chalov,
Phys. Rev. {\bf D 68} 094008 (2003). 
\bibitem{j-parc} ``The Joint Project for High-Intensity Proton
Accelerators'', KEK Report 99-4 (1999); http://j-parc.jp/
\end{thebibliography}
\end{document}